\documentclass[journal,article,submit,pdftex,moreauthors]{mdpi} 
\usepackage{amsmath,amssymb,amsfonts,amsthm}
\newcommand{\bs}{\boldsymbol}
\newcommand{\nn}{\nonumber}
\firstpage{1} 
\pubvolume{1}
\issuenum{1}
\articlenumber{0}
\pubyear{2023}
\copyrightyear{2023}
\datereceived{ } 
\daterevised{ } 
\dateaccepted{ } 
\datepublished{ } 
\hreflink{https://doi.org/} 

\Title{Recent Developments in Quarkonium as an Open Quantum System in Quark-Gluon Plasma}

\TitleCitation{Recent Developments in Quarkonium as an Open Quantum System in Quark-Gluon Plasma}

\Author{Bruno Scheihing Hitschfeld$^{1}$\orcidA{} and Xiaojun Yao$^{2}$*\orcidB{}}

\AuthorNames{Bruno Scheihing Hitschfeld and Xiaojun Yao}

\AuthorCitation{Bruno Scheihing Hitschfeld; Xiaojun Yao}

\address{%
$^{1}$ \quad Center for Theoretical Physics, Massachusetts Institute of Technology, Cambridge, MA  02139, USA; bscheihi@mit.edu\\
$^{2}$ \quad InQubator for Quantum Simulation, Department of Physics, University of Washington, Seattle, WA 98195, USA; xjyao@uw.edu}

\corres{Correspondence: xjyao@uw.edu}

\abstract{We review recent progress in understanding quarkonium dynamics inside the quark-gluon plasma as an open quantum system with a focus on the definition and nonperturbative calculations of relevant transport coefficients and generalized gluon distributions.}

\keyword{Quarkonium; quark-gluon plasma; heavy ion collision; open quantum system; effective field theory; AdS/CFT correspondence.\\ \\ \textit{Report number:} MIT-CTP/5663, IQuS@UW-21-068} 

\begin{document}




\vspace{-.45cm}

\section{Introduction}

Quarkonia, bound states of heavy quark-antiquark ($Q\bar{Q}$) pairs, have played a unique role in probing the properties of the quark-gluon plasma (QGP). Early studies showed that the attractive potential between the heavy quark-antiquark ($Q\bar{Q}$) pair inside a quarkonium state could get color-screened by the deconfined nuclear matter, i.e., the QGP, leading to a (static) melting of the quarkonium bound state. The absence of quarkonium bound states inside a high temperature QGP can result in a suppression of the quarkonium yield in heavy-ion collisions compared to that in proton-proton collisions where the QGP is not expected to form. Therefore, quarkonium suppression is considered to be strong evidence for the existence of the deconfined QGP. Furthermore, since the occurrence of melting depends on an interplay between the quarkonium binding energy, which is inversely proportional to its size, and Debye screening, which is directly related to the QGP temperature, quarkonia of different binding energies and thus different sizes are expected to dissociate at different temperatures, or experience different levels of suppression at the same temperature. As such, quarkonium can be utilized as a "thermometer" for the QGP. 

While the static color-screening picture is simple and appealing, it has been realized more recently that other hot medium effects are also important and need to be considered when interpreting quarkonium suppression data in heavy-ion collisions. For example, dynamical processes, such as inelastic scatterings between partons from the QGP and quarkonia in a weak coupling picture, can lead to quarkonium dissociation and recombination. Understanding these dynamical medium effects in an expanding QGP fireball requires a real-time description. Furthermore, the strong coupling nature of the QGP created in current collision experiments has made such a description even more challenging due to the lack of a nonperturbative generalization of the weak coupling picture driven by inelastic scatterings mentioned above.

In the last few years, much progress has been achieved by using the open quantum system (OQS) framework and nonrelativistic effective field theories (NREFT) of QCD (see recent reviews~\cite{Rothkopf:2019ipj,Yao:2021lus}). Important new theoretical developments include: (1) the understanding of quarkonium dissociation as a result of quantum decoherence of the quarkonium wavefunction; (2) the establishment of the deep relation between quarkonium dissociation and recombination in both quantum and classical transport approaches; and (3) the construction and calculations of novel transport coefficients for a $Q\bar{Q}$ pair entangled in color. Here we will review these achievements by discussing in-medium quarkonium time evolution equations in three temperature ranges, all of which together cover the whole temperature history of the QGP created in current heavy ion collisions. We will emphasize how the QGP properties relevant for quarkonium dynamics are encoded into the evolution equations nonperturbatively.

\section{Quarkonium dynamics inside QGP: OQS and NREFT}
Quarkonium in vacuum is characterized by a hierarchy of three energy scales: $M\gg Mv \gg Mv^2$, where $M$ is the heavy quark mass, and $v$ the typical relative velocity between a bound $Q\bar{Q}$ pair. The typical size of a quarkonium state is given by $r\sim \frac{1}{Mv}$ and its typical binding energy is given by $Mv^2$. Inside the QGP, depending on where the temperature $T$ fits the hierarchy, we can develop different effective descriptions. Before detailed discussions, we want to emphasize two things. First, we assume the QGP is strongly coupled so we do not distinguish the scale $T$ and the scale $gT$ which is roughly the Debye mass. Second, the essence of the different effective descriptions is a choice of the computational basis. Under a given hierarchy of energy scales, certain choice of the basis turns out to be more useful in the sense that a low truncation of the basis can describe most of the important physics.

\subsection{$M\gg T \gtrsim Mv$}
Under this hierarchy, the $Q\bar{Q}$ potential is significantly screened. As a result, the $Q\bar{Q}$ pair behaves more like two individual heavy quarks that interact weakly with each other. The most appropriate effective description is obtained by combining the OQS with the nonrelativistic QCD (NRQCD). A Lindblad equation in the so-called quantum Brownian motion limit can be derived to describe the dynamics of the $Q\bar{Q}$ pair~\cite{Akamatsu:2014qsa}, concretely
\begin{align}
\label{eqn:lindblad_nrqcd}
\frac{{\rm d}\rho_S(t)}{{\rm d} t} &=  -i \big[ H_S+\Delta H_S, \rho_S(t)\big] + \frac{1}{N_c^2-1}\int\frac{{\rm d}^3q}{(2\pi)^3} \, D^{>}(q_0=0,{\bs q}) \nn\\
&\qquad\qquad\qquad \times \big( \widetilde{O}^a({\bs q}) \rho_S(t) \widetilde{O}^{a\dagger}({\bs q})  - \frac{1}{2} \{  \widetilde{O}^{a\dagger}({\bs q}) \widetilde{O}^a({\bs q}) , \rho_S(t) \} \big) \,.
\end{align}
We focus on the non-unitary part consisting of the Lindblad operator and the environment correlator
\begin{align}
\widetilde{O}^a({\bs q}) &= e^{\frac{i}{2}{\bs q}\cdot \hat{\bs x}_Q} \Big( 1-\frac{{\bs q}\cdot \hat{\bs p}_Q }{4MT}  \Big) e^{\frac{i}{2}{\bs q}\cdot \hat{\bs x}_Q} T_F^a - e^{\frac{i}{2}{\bs q}\cdot \hat{\bs x}_{\bar{Q}}} \Big( 1-\frac{{\bs q}\cdot \hat{\bs p}_{\bar{Q}} }{4MT}  \Big) e^{\frac{i}{2}{\bs q}\cdot \hat{\bs x}_{\bar{Q}}} T_F^{*a} \\
D^>(t,{\bs x}) &= g^2 \big\langle A_0^a(t,{\bs x}) A_0^a(0,{\bs 0}) \big\rangle_T  = g^2{\rm Tr}_E \big( A_0^a(t,{\bs x}) A_0^a(0,{\bs 0}) \rho_E \big) \,,
\end{align}
where $\hat{\bs x}$, $\hat{\bs p}$ are one-particle position and momentum operators and $\rho_E$ is the environment density matrix at thermal equilibrium.
The $\frac{1}{T}$ term inside $\widetilde{O}^a({\bs q})$ originates from a high temperature expansion (i.e., $\frac{H_S}{T}$) in the quantum Brownian motion limit and accounts for dissipative effects~\cite{Kajimoto:2017rel}. This term is necessary for an approximate thermalization of the $Q\bar{Q}$ system and it slows down the quarkonium dissociation process in the early stage of the time evolution. The Lindblad operator $\widetilde{O}^a({\bs q})$ generates an extra phase that is position dependent: the wavefunction at different positions evolves with different phases during the time evolution. As a result, the wavefunction loses coherence overtime. Both dissociation and recombination can be thought of as a consequence of the quantum decoherence. 

The classical limit of Eq.~\eqref{eqn:lindblad_nrqcd} corresponds to a Fokker-Planck equation, which is equivalent to a Langevin equation. This explains why this limit is called the quantum Brownian motion limit. Under the gradient expansion in the classical limit, the $D^>(t,{\bs x})$ correlator encodes the same information as $\langle E_i^a(t,{\bs x}) E_i^a(0,{\bs 0})\rangle_T$, which at zero frequency corresponds to the leading order expansion in powers of $g$ of the heavy quark diffusion coefficient $\kappa_{\rm fund}$
\begin{align}
\kappa_{\rm fund} = \frac{g^2}{3N_c} {\rm Re} \! \int \! {\rm d} t  \big\langle {\rm Tr}_{\rm c}[ U(-\infty,t) E_i(t) U(t,0) E_i(0) U(0, -\infty)  ] \big\rangle_{T,Q} \,,
\end{align}
where the subscript $Q$ indicates the effect of the heavy quark on the thermal density matrix. 

In short, the intuitive physical picture in this hierarchy is that the $Q$ and $\bar{Q}$ diffuse independently while interacting via a weak potential.

\subsection{$M\gg Mv \gg T \gg Mv^2$}
In this hierarchy, the interaction potential between the $Q\bar{Q}$ pair is no longer negligible. One can further expand NRQCD in powers of $r\sim\frac{1}{Mv}$ and obtain potential NRQCD (pNRQCD) as the effective description. The hierarchy $T\gg Mv^2$ allows an expansion of $\frac{H_S}{T}$, which also corresponds to the quantum Brownian motion limit discussed above. The resulting Lindblad equation is~\cite{Brambilla:2016wgg}:
\begin{align}
\label{eqn:lindblad}
\frac{{\rm d} \rho_S(t)}{{\rm d} t} =  -i\big[ H_S + \gamma_{\rm adj} \Delta h_S,\, \rho_S(t) \big] +  \kappa_{\rm adj} \big( L_{\alpha i} \rho_S(t) L^\dagger_{\alpha i} - \frac{1}{2}\{  L^\dagger_{\alpha i}L_{\alpha i},\, \rho_S(t)\} \big)\,.
\end{align}
The quarkonium transport coefficients ($\kappa_{\rm adj}$, $\gamma_{\rm adj}$ are real) are defined nonperturbatively as
\begin{align}
\label{eqn:adj}
\kappa_{\rm adj} + i\gamma_{\rm adj} &\equiv \frac{ g^2 T_F}{3 N_c}  \int {\rm d} t\, \big\langle \mathcal{T} E^a_i(t) W^{ab}(t,0) E^b_i(0) \big\rangle_T \,,
\end{align}
where $\mathcal{T}$ denotes time ordering and $W^{ab}(t,0)$ is a straight timelike adjoint Wilson line.

The correlator involved in the quarkonium case is different from that in the single heavy quark case in terms of operator ordering~\cite{Scheihing-Hitschfeld:2022xqx}. Recently, we have made progress towards a nonperturbative understanding of these transport coefficients. We performed an AdS/CFT calculation and showed $\kappa_{\rm adj}=\gamma_{\rm adj}=0$ in a strongly coupled $\mathcal{N}=4$ supersymmetric Yang-Mills plasma~\cite{Nijs:2023dks}. We also discussed how to extract these transport coefficients from lattice QCD calculations of an Euclidean correlator~\cite{Scheihing-Hitschfeld:2023tuz}
\begin{align}
G_{\rm adj}(\tau) = - \frac{g^2 T_F }{3 N_c} \big \langle E_i^a(\tau) W^{ab}(\tau,0) E_i^b(0) \big\rangle_T =  \int_{-\infty}^{+\infty} \frac{{\rm d} \omega}{2\pi} \frac{\exp \big( \omega ( \frac{1}{2T} - \tau ) \big)}{2\sinh \big( \frac{\omega}{2T} \big) } \rho_{\rm adj}^{++}(\omega) \,,
\end{align}
where the spectral function $\rho_{\rm adj}^{++}(\omega)$ is non-odd in $\omega$. Our analyses determined
\begin{align}
\label{eqn:extraction}
\kappa_{\rm adj} &= \lim_{\omega\to0} \frac{T}{2\omega} \left [\rho_{\rm adj}^{++}(\omega) - \rho_{\rm adj}^{++}(-\omega) \right] \, , \\
\gamma_{\rm adj} &= - \int_0^\beta {\rm d} \tau \, G_{\rm adj}(\tau) - \frac{1}{2\pi} \int_{-\infty}^{+\infty} \!\!\! {\rm d} \omega  \frac{1 + 2n_B(|\omega|)}{|\omega|} \rho_{\rm adj}^{++}(\omega) \,. 
\end{align}

\subsection{$M\gg Mv \gg Mv^2 \gtrsim T$}
In this hierarchy, pNRQCD is applicable but the quantum Brownian motion limit breaks down due to the fact that $T$ is on the same order as or smaller than $Mv^2$, i.e., $Mv^2 \gtrsim T$. The quantum optical limit of OQS fits here and the basis of bound state and scattering state wavefunctions is a better choice than the position basis used in the two previous cases. At present, no well-defined Lindblad equation has been found for this case. However, a semiclassical limit of the Markovian quantum evolution is well-defined~\cite{Yao:2021lus}, given in terms of a Boltzmann (rate) equation~\cite{Yao:2018nmy}
\begin{align}
\label{eqn:rate}
\frac{{\rm d} n_b(t,{\bs x})}{{\rm d} t} = -\Gamma\, n_b(t,{\bs x}) + F(t,{\bs x}) \,,
\end{align}
where the dissociation rate and the recombiantion contribution are 
\begin{align}
\label{eqn:disso}
\Gamma &=  \int \frac{{\rm d}^3p_{\mathrm{rel}}}{(2\pi)^3} 
| \langle \psi_b | {\bs r} | \Psi_{{\bs p}_\mathrm{rel}} \rangle |^2 [g^{++}_E]^{>}(-\Delta E) \\
\label{eqn:reco}
F &=  \int \frac{{\rm d}^3p_{\mathrm{cm}}}{(2\pi)^3}  \frac{{\rm d}^3p_{\mathrm{rel}}}{(2\pi)^3} 
| \langle \psi_b | {\bs r} | \Psi_{{\bs p}_\mathrm{rel}} \rangle |^2 
[g^{--}_E]^{>}(\Delta E) f_{Q\bar{Q}}(t, {\bs x}, {\bs p}_{\mathrm{cm}}, {\bs x}_{\rm rel}=0, {\bs p}_{\mathrm{rel}}) \,,  
\end{align}
where $\Delta E>0$ is the energy gap between the bound state $\psi_b$ and the scattering state $\Psi_{{\bs p}_{\rm rel}}$. 

The generalized gluon distribution (GGD) for dissociation is defined in terms of the same chromoelectric field correlator introduced above~\cite{Yao:2020eqy,Binder:2021otw}
\begin{align}
[g_{\rm adj}^{++}]^>(\omega) \equiv \frac{g^2 T_F }{3 N_c} \int\frac{{\rm d} \omega}{2\pi} e^{i\omega t} \langle E_i^a(t) {W}^{ab}(t,0) E_i^b(0) \rangle_T \,.
\end{align}
The GGD $[g_{\rm adj}^{--}]^>$ for recombination is related to $[g_{\rm adj}^{++}]^>$ via a generalized KMS relation~\cite{Binder:2021otw}
\begin{align}
\label{eqn:kms}
[g_{\rm adj}^{++}]^>(\omega) = e^{\omega/T} [g_{\rm adj}^{--}]^>(-\omega) \,,
\end{align}
which is necessary for the system to reach detailed balance between dissociation and recombination. It has been shown that $\kappa_{\rm adj} = [g_{\rm adj}^{++}]^>(\omega=0)$~\cite{Scheihing-Hitschfeld:2022xqx}.

We calculated the GGDs perturbatively to next-to-leading-order accuracy at finite temperature~\cite{Binder:2021otw}. We also calculated them in a strongly-coupled $\mathcal{N}=4$ supersymmetric Yang-Mills plasma and found $g_{\rm adj}^{++}(\omega)=0$ when $\omega \leq 0$, which indicates the Boltzmann (rate) equation~\eqref{eqn:rate} becomes trivial~\cite{Nijs:2023dbc}. This striking result in the strong coupling limit urged for the development of non-Markovian descriptions of in-medium quarkonium dynamics~\cite{Nijs:2023bok}. These descriptions will be crucial to understand quarkonium production in current heavy ion collision experiments, where the QGP created is strongly coupled.

\section{Future directions}
As we review the recent progress, it becomes clear that it is urgent to calculate the quarkonium transport coefficients and GGDs nonperturbatively in QCD. At the same time, developing non-Markovian descriptions of the time evolution of quarkonium in the QGP is paramount to be able to account for the effects of a strongly coupled plasma.

\vspace{6pt}

\funding{This work is supported by the U.S.~Department of Energy, Office of Science, Office of Nuclear Physics under grant Contract Number DE-SC0011090\@. X.Y. is supported by the U.S. Department of Energy, Office of Science, Office of Nuclear Physics, InQubator for Quantum Simulation (IQuS) (https://iqus.uw.edu) under Award Number DOE (NP) Award DE-SC0020970 via the program on Quantum Horizons: QIS Research and Innovation for Nuclear Science.}

\conflictsofinterest{The authors declare no conflict of interest.
}

\begin{adjustwidth}{-\extralength}{0cm}

\reftitle{References}

\end{adjustwidth}

\begin{thebibliography}{999}
\bibitem{Rothkopf:2019ipj}
A.~Rothkopf,
Phys. Rept. \textbf{858}, 1-117 (2020);
Y.~Akamatsu,
Prog. Part. Nucl. Phys. \textbf{123}, 103932 (2022);
R.~Sharma,
Eur. Phys. J. ST \textbf{230}, no.3, 697-718 (2021).

\bibitem{Yao:2021lus}
X.~Yao,
Int. J. Mod. Phys. A \textbf{36}, no.20, 2130010 (2021).

\bibitem{Akamatsu:2014qsa}
Y.~Akamatsu,
Phys. Rev. D \textbf{91}, no.5, 056002 (2015);
J.~P.~Blaizot and M.~A.~Escobedo,
JHEP \textbf{06}, 034 (2018).

\bibitem{Kajimoto:2017rel}
S.~Kajimoto, Y.~Akamatsu, M.~Asakawa and A.~Rothkopf,
Phys. Rev. D \textbf{97}, no.1, 014003 (2018);
Y.~Akamatsu, M.~Asakawa, S.~Kajimoto and A.~Rothkopf,
JHEP \textbf{07}, 029 (2018);
T.~Miura, Y.~Akamatsu, M.~Asakawa and A.~Rothkopf,
Phys. Rev. D \textbf{101}, no.3, 034011 (2020).

\bibitem{Brambilla:2016wgg}
N.~Brambilla, M.~A.~Escobedo, J.~Soto and A.~Vairo,
Phys. Rev. D \textbf{96}, no.3, 034021 (2017);
N.~Brambilla, M.~A.~Escobedo, J.~Soto and A.~Vairo,
Phys. Rev. D \textbf{97}, no.7, 074009 (2018);
N.~Brambilla, M.~\'A.~Escobedo, A.~Islam, M.~Strickland, A.~Tiwari, A.~Vairo and P.~Vander Griend,
Phys. Rev. D \textbf{108}, no.1, L011502 (2023).

\bibitem{Scheihing-Hitschfeld:2022xqx}
B.~Scheihing-Hitschfeld and X.~Yao,
Phys. Rev. Lett. \textbf{130}, no.5, 5 (2023).

\bibitem{Nijs:2023dks}
G.~Nijs, B.~Scheihing-Hitschfeld and X.~Yao,
JHEP \textbf{06}, 007 (2023).

\bibitem{Scheihing-Hitschfeld:2023tuz}
B.~Scheihing-Hitschfeld and X.~Yao,
Phys. Rev. D \textbf{108}, no.5, 054024 (2023).

\bibitem{Yao:2018nmy}
X.~Yao and T.~Mehen,
Phys. Rev. D \textbf{99}, no.9, 096028 (2019);
X.~Yao, W.~Ke, Y.~Xu, S.~A.~Bass and B.~M\"uller,
JHEP \textbf{01}, 046 (2021);
X.~Yao and B.~M\"uller,
Phys. Rev. C \textbf{97}, no.1, 014908 (2018);
Phys. Rev. D \textbf{97}, no.7, 074003 (2018);
Phys. Rev. D \textbf{100}, no.1, 014008 (2019).

\bibitem{Yao:2020eqy}
X.~Yao and T.~Mehen,
JHEP \textbf{02}, 062 (2021).

\bibitem{Binder:2021otw}
T.~Binder, K.~Mukaida, B.~Scheihing-Hitschfeld and X.~Yao,
JHEP \textbf{01}, 137 (2022).

\bibitem{Nijs:2023dbc}
G.~Nijs, B.~Scheihing-Hitschfeld and X.~Yao,
[arXiv:2310.09325 [hep-ph]].

\bibitem{Nijs:2023bok}
G.~Nijs, B.~Scheihing-Hitschfeld and X.~Yao,
[arXiv:2312.12307 [hep-ph]].

\end{thebibliography}
\end{document}